\documentclass[aps,amsmath,amssymb,showpacs,pre]{revtex4-1}
\newcommand{\be}{\begin{equation}}
\newcommand{\ee}{\end{equation}}
\newcommand{\bea}{\begin{eqnarray}}
\newcommand{\eea}{\end{eqnarray}}
\usepackage{epsfig}
\usepackage{bm}
\usepackage{latexsym}
\usepackage{graphicx}
\usepackage{comment}

\usepackage{graphicx,color}
 \graphicspath{{./}{./figures/}}
\usepackage{ams math}
\usepackage{enumerate}
\usepackage{mathbbol}
\usepackage{amsfonts}
\usepackage{natbib}

\usepackage{bm}
\usepackage{hyperref}

\usepackage{color}

\begin{document}

\title{The effect of driving on model C interfaces}
\author{David S. Dean$^{1,2}$, Paul Gersberg$^{1,3}$ and P.C.W. Holdsworth$^{3}$}
\affiliation{(1) Univ. Bordeaux, CNRS, LOMA, UMR 5798, F-33400 Talence, France}
\affiliation{(2) School of Physical Sciences and Kavli Institute for Theoretical Sciences, University of Chinese Academy of Sciences, Beijing 100049, China}
\affiliation{(3) Universit\'e de Lyon, ENS de Lyon, Universit\'e Claude Bernard
CNRS, Laboratoire de Physique, F-69342 Lyon, France}
\bibliographystyle{apsrev}
\begin{abstract}{We consider the effect of uniform driving on the interface between two phases which are described by model C dynamics. The non-driven system has a classical Gaussian interface described by capillary wave theory. The model under driving retains Gaussian statistics but the interface statistics are modified by driving, notably the height fluctuations are suppressed and  the correlation length of the fluctuations is increased.
The model we introduce can also be used as a model for the effect of activity on interface dynamics. }
\end{abstract}
\maketitle
\section{Introduction}
One of the most natural ways of creating a non-equilibrium steady state is by applying external  driving forces. Driving arises naturally in sedimenting systems due to gravity, in systems with free charges under the action of an electric field and also  due to the radiation pressure exerted by a laser. Experiments where a phase separated colloidal system is sheared parallel to the interface show that  driving due to shear tends to suppress surface fluctuations \cite{derks2006},
and similar results are found where Ising models are numerically sheared 
\cite{smith2008,smith2010}. These results are somewhat surprising, for instance they are contrary to the  observation that wind generates waves on the ocean. One may think that the precise nature of the driving plays a role, for instance uniformly driving a system may be intrinsically different to applying a shear field which is manifestly nonuniform. In this paper we investigate analytically the effect of uniform driving on a simple interface model. We find that the effect of this type of  driving is also to reduce surface fluctuations.

Constructing a continuum model which is analytically tractable and is also 
affected by uniform driving is straightforward but contains some subtleties. 
In a continuum system it is clear that uniform driving can only move a system away from equilibrium when the driving acts differently on different particle types. For instance, consider  a system of identical interacting Brownian particles driven by a uniform  force. The force will induce the same average velocity on all the particles, consequently, in the frame moving with this average velocity, we will recover the unmodified equilibrium state. However, when multiple particle types are present, the mean velocity induced on different species are different and no Galilean transformation is possible. Perhaps the first such study of this phenomenon was due to Onsager \cite{hem1996}, who studied the conductivity of electrolytes and in doing so showed how the correlation functions in the steady state were modified by the electric field. 
Recently there have been many studies of driven multi-particle Brownian systems \cite{netz2003,dzub2002,chak2003,chak2004,lowe2009,glan2012, klym2016}, including the electrolyte problem,  and rich new physics has been found, even in the case of purely Gaussian theories \cite{dem2016,pon2017} based on stochastic density functional theory \cite{dean1996}. 

The dynamics of discrete particle systems is however affected by uniform driving of identical particles. The study of driven lattice gases has revealed a wide range of intriguing physical phenomena and indeed shown how driving can even lead to phase separation \cite{katz1984,zia1991,leun1993,schm1995,schm1998}. The discrete nature of the dynamics of these systems, both in space and time, means that no Galilean transformation to an equilibrium state exists. Analytical studies of these systems require a phase ordering kinetics description in terms of a continuum order parameter. In order to break Galilean invariance the local mobility of the particles can be taken to be dependent on the local order parameter, this is then sufficient to induce non-trivial steady states under driving \cite{katz1984,leun1993,schm1995,schm1998}. Interfaces between the separated phases in uniformly driven systems have non capillary behaviors which are, even today, not fully understood \cite{leun1993}.
Taking random driving in a given direction also leads to non-equilibrium steady states, if the noise is Gaussian and white, the fluctuation dissipation theorem is violated and novel interface fluctuations are induced which, again, are  not of  the capillary type \cite{zia1991}. 

Driving can also be deterministic but space dependent, for instance if one considers applied shear flows, the spatial dependence of the flow means no Galilean transformation to an equilibrium steady state is possible and this therefore leads to non-equilibrium steady states. The effect of shear on interfaces in these type of systems yields interface equations of the stochastic Burgers type and the statistics are no thus longer Gaussian due to the presence of nonlinearities \cite{bray2001,bray2002,smith2008,smith2010,thie2010,thie2014}. 

In this paper we analyse what is known, in the classification of Hohenberg and Halperin \cite{hohe1977}, as model C type dynamics for two fields, one with conserved model B type dynamics, which is in addition convected at a uniform velocity to mimic driving. We refer to this first field as the colloid field.
This colloid field is coupled to an additional field which undergoes model A non-conserved dynamics and which is not subjected to the driving. The model A field can be thought of a passive solvent and its coupling to the model B field is chosen in such a way that it has no influence on the non-driven equilibrium steady state. We then derive the effective dynamics between two separated low temperature phases by using a
method introduced in \cite{bray2001,bray2002} for the study of interfaces under shear flow. This method yields a Gaussian theory for the interface statistics and driving introduces interesting new physics, notably we find that the effective surface tension of the system is increased but also the correlation length of interface fluctuations (due to an effective gravitational term) are increased. These observations are in qualitative agreement with experimental results on sheared low tension interfaces in phase separated colloidal systems \cite{derks2006}. In this experimental system the interface fluctuations were also found to be well described by Gaussian statistics and this is our principal motivation for studying theories which remain Gaussian but are  modified by driving. While the long wavelength theory we find is of a capillary type, we also find new, higher derivative terms, which  are generated in the spectrum of the height fluctuations. 

As an aside, we also show how the model introduced here can be used to analyse the effect of activity on the dynamics of the surface between two phases of active colloids. The activity is implemented by taking a different temperature for the colloid and solvent fields, this difference in temperatures leads to significantly modified surface statistics which again develop dependencies on static and dynamical variables of the model which otherwise remain hidden for the equilibrium version of the problem.

\section{The underling two field  model}
We consider a coarse grained model for two scalar fields $\psi$ and $\phi$ with Hamiltonian
\begin{equation}
H[\psi,\phi] = H_1[\psi] +H_2[\psi,\phi]
\end{equation}
The Hamiltonian $H_1$ is of the classic Landau-Ginzburg form
\begin{equation}
H_1[\psi]=\int d{\bf x}\left[\frac{\kappa}{2}[\nabla\psi({\bf x})]^2 + V(\psi({\bf x}))
- gz \psi({\bf x})\right].
\end{equation}
The last term represents the energy due to a gravitational field and will introduce a finite correlation length in the fluctuations between the two phases. We assume that the above Hamiltonian has two stable phases with average concentrations of the field $\phi({\bf x})$ given by the constant values $\psi_1$ and $\psi_2$, the difference between the order parameter in  the two different phases is denoted by 
by $\Delta\psi= \psi_2 -\psi_1>0$. This means that we find the phase $1$ as $z\to\infty$ and the phase $2$ as $z\to-\infty$. The term $H_2$ is taken to be a simple quadratic coupling between the fields
\begin{equation}
H_2 =\int d{\bf x} \frac{\lambda}{2}(1-\psi({\bf x})-\phi({\bf x}))^2,
\end{equation}
this is an approximative conservation law of total volume fraction of the phases. The field $\phi$ can be though of as the local volume fraction of the solvent in a colloidal system. However the presence of this solvent field does not change the effective equilibrium statistical mechanics of the colloid field $\psi$ as the partition function can be written as 
\begin{equation}
Z = \int d[\phi]d[\psi]\exp(-\beta H_1[\psi]- \beta H_2[\psi,\phi]) = CZ_{eff},
\end{equation}
where $Z_{eff}$ is the effective partition function for the field $\psi$, after we have integrated out the degrees of freedom corresponding to the field $\phi$,
and $C$ is a constant term resulting from this integration. The effective partition function is thus simply given by
\begin{equation}
Z_{eff} = \int d[\psi]\exp(-\beta H_1[\psi]),
\end{equation}
and, as stated above, we see that the field $\phi$ thus has no effect on the equilibrium statistical mechanics of the field $\psi$.

We now consider the dynamics of the fields. We take local diffusive model B dynamics for the field $\psi$ and non-conserved model A dynamics for the field $\phi$
\begin{eqnarray}
\frac{\partial \psi({\bf x},t)}{\partial t} +{\bf v}\cdot { \nabla}\psi({\bf x},t)&=& D\nabla^2\frac{\delta H}{\delta \psi({\bf x})}+ \sqrt{2D T}\nabla \cdot {\bm \eta}_1({\bf x},t) \\
\frac{\partial \phi({\bf x},t)}{\partial t} &=& -\alpha\frac{\delta H}{\delta \phi({\bf x})}+ \sqrt{2\alpha T}{ \eta}_2({\bf x},t).
\end{eqnarray}
The first equation corresponds to standard model B dynamics but with an advection term by a constant velocity field $\bf v$. The second equation has no advection term and is simple model A dynamics. In principle we can also treat the case where the dynamics of the field $\phi$ is also diffusive and thus of model $B$ type, the analysis given here can be extended to this case but the analysis of the resulting equations is considerably more complicated. The use of model A dynamics for the solvent is justified by assuming that its dynamics is faster than that of the colloids and that the volume fraction can vary due to local conformational changes rather than  diffusive transport.

The noise terms above 
are uncorrelated and Gaussian with zero mean, their correlation functions are given by
\begin{eqnarray}
\langle \eta_{1i}({\bf x},t) \eta_{1j}({\bf x}',t)\rangle&=& \delta_{ij}\delta(t-t') \delta({\bf x}-{\bf x}') \\
\langle \eta_{2}({\bf x},t) \eta_{2}({\bf x}',t)\rangle&=& \delta(t-t') \delta({\bf x}-{\bf x}') ,
\end{eqnarray}
and $T$ is the temperature in units where $k_B=1$.
These dynamical equations  are thus explicitly given by
\begin{equation}
\frac{\partial \psi({\bf x},t)}{\partial t} +{\bf v}\cdot { \nabla}\psi({\bf x},t)= D\nabla^2[\frac{\delta H_1}{\delta \psi({\bf x})}+\lambda(\phi({\bf x},t) + \psi({\bf x},t))]+ \sqrt{2D T}\nabla \cdot {\boldsymbol \eta}_1({\bf x},t)
\end{equation}
and
\begin{equation}
\frac{\partial \phi({\bf x},t)}{\partial t} = -\alpha\lambda[\phi({\bf x},t) + \psi({\bf x},t)]+ \sqrt{2\alpha T}{ \eta}_2({\bf x},t).
\end{equation}
Taking the temporal Fourier transform, defined with the convention
\begin{equation}
\tilde F({\bf x}, \omega) = \int_{-\infty}^\infty dt \exp(-i\omega t)F({\bf x}, t),
\end{equation}
we can eliminate the field $\tilde \phi$ which is given by
\begin{equation}
\tilde \phi({\bf x},\omega) = \frac{-\alpha\lambda \tilde \psi({\bf x},\omega)+\sqrt{2\alpha T}\tilde \eta_2({\bf x},\omega)}{i\omega +\alpha \lambda},
\end{equation}
this then gives the closed equation for $\tilde \psi$:
\begin{equation}
\left[1-\frac{\lambda D \nabla^2}{i\omega+\alpha\lambda}\right]i\omega \tilde\psi({\bf x}, \omega) +{\bf v}\cdot\nabla\tilde\psi({\bf x}, \omega)
= D\nabla^2 \tilde \mu({\bf x},\omega) +  \tilde \zeta({\bf x},\omega),
\end{equation}
where 
\begin{equation}
\mu({\bf x},t)=\frac{\delta H_1}{\delta \psi({\bf x},t)}
\end{equation}
is the effective chemical potential associated with the field $\psi$ and the noise term is given by
\begin{equation}
\tilde \zeta({\bf x},\omega) = \frac{\sqrt{2\alpha T}D\lambda}{i\omega + \alpha\lambda}\nabla^2\tilde \eta_2({\bf x},\omega) +
\sqrt{2DT}\nabla\cdot\tilde {\bm \eta}_1({\bf x},\omega).
\end{equation}
Inverting the temporal Fourier transform then gives the effective evolution equation
\begin{equation}
\frac{\partial \psi({\bf x},t)}{\partial t} -\lambda D\nabla^2\int_{-\infty}^t dt'
\exp(-\alpha\lambda(t-t')) \frac{\partial \psi({\bf x},t')}{\partial t}+{\bf v}\cdot\nabla\psi({\bf x}, t)=D\nabla^2  \mu({\bf x},t') +  \zeta({\bf x},t).\label{dyn1}
\end{equation}
\section{Effective interface dynamics}
We now follow the method of \cite{bray2001,bray2002} to derive the dynamical equation  for the interface between the two phases. It is assumed that the driving is in the ${\bf r}=(x,y)$ plane and that the system varies from phase $1$ to phase $2$ in the $z$ direction. The dynamical evolution for the field $\psi$ in Eq. (\ref{dyn1}) is first written as
\begin{equation}
\nabla^{-2}\left[\frac{\partial \psi({\bf x},t)}{\partial t}+{\bf v}\cdot\nabla\psi({\bf x}, t)\right] -\lambda D\int_{-\infty}^t dt'
\exp(-\alpha\lambda(t-t')) \frac{\partial \psi({\bf x},t')}{\partial t'}=D  \mu({\bf x},t') + \nabla^{-2} \zeta({\bf x},t).\label{eqpsi}
\end{equation}
We now assume that the field $\psi$ can be written in the form
\begin{equation}
\psi({\bf x},t) = f(z-h({\bf r},t)),
\end{equation}
and $f(z)\to \psi_2$ as $z\to -\infty$ and $f(z)\to \psi_2$ as  $z\to \infty$.
We now note the following results
\begin{eqnarray}
\frac{\partial f(z-h({\bf r},t))}{\partial t} &=& -f'(z-h({\bf r},t))\frac{h({\bf r},t)}{\partial t}\\
\nabla f(z-h({\bf r},t) )&=& [{\bf e}_z -\nabla h({\bf r},t)]f'(z-h({\bf r},t))]\\
\nabla^2 f(z-h({\bf r},t)) &=& f''(z-h({\bf r},t)[1 + [\nabla h({\bf r},t)]^2] -\nabla^2 h({\bf r},t)f'(z-h({\bf r},t)),
\end{eqnarray}
and thus we find
\begin{equation}
\mu({\bf x},t)= -\kappa\left(f''(z-h({\bf r},t)[1 + \nabla^2 h({\bf r},t)] -\nabla^2 h({\bf r},t)f'(z-h({\bf r},t))\right) + V'(f(z-h({\bf r},t)) - gz .
\end{equation}
Multiplying both sides of the above by $f'(z-h({\bf r},t))$ yields
\begin{eqnarray}
&&f'(z-h({\bf r},t))\mu({\bf x},t)=\nonumber \\
 &&-\kappa\left(f'(z-h({\bf r},t)f''(z-h({\bf r},t)[1 + \nabla^2 h({\bf r},t)] -\nabla^2 h({\bf r},t)f'(z-h({\bf r},t))^2\right) + V'(f(z-h({\bf r},t))f'(z-h({\bf r},t))\nonumber \\
 &&- gz f'(z-h({\bf r},t)) \nonumber
\end{eqnarray}
and then integrating over $z$ we obtain
\begin{eqnarray}
\int_{-\infty}^\infty dz f'(z-h({\bf r},t)\mu({\bf x},t)&=& \kappa \nabla^2 h({\bf r},t)\int_{-\infty}^\infty dz\ f'(z-h({\bf r},t))^2 - \int_{-\infty}^\infty dz gz f'(z-h({\bf r},t))\nonumber \\&=&
\kappa\nabla^2 h({\bf r},t)\int_{-\infty}^\infty dz'\ f'(z')^2 - \int_{-\infty}^\infty dz' g(z' +h({\bf r},t)) f'(z')\nonumber \\
&=& \kappa\nabla^2 h({\bf r},t)\int_{-\infty}^\infty dz' \ f'(z')^2 -\Delta\psi g h({\bf r},t).
\end{eqnarray}
In the above we have assumed that $\int_{-\infty}^\infty dz' z'f'(z')=0$ by symmetry (this is also consistent with the approximation made later on in Eq. (\ref{eqdelta})). Furthermore one can show that \cite{bray2001,bray2002}
\begin{equation}
\kappa\int_{-\infty}^\infty dz' \ f'(z')^2 = \sigma,\label{mfsig}
\end{equation}
where $\sigma$ is the mean-field equilibrium Cahn-Hilliard estimate of the surface tension, obtained by  assuming that $f(z)=\psi_{MF}(z)$ is the equilibrium mean field profile of the field 
$\psi$. We thus find
\begin{equation}
\int_{-\infty}^\infty dz f'(z-h({\bf r},t)\mu({\bf x},t) = \sigma[\nabla^2 h({\bf r},t)-m^2 h({\bf r},t)]
\end{equation}
where $m^2 = \Delta\psi g /\sigma$. We now carry out the same operation on the left hand side of Eq. (\ref{eqpsi}). First we have
\begin{eqnarray}
\nabla^{-2}\frac{\partial \psi({\bf x},t)}{\partial t}&+&{\bf v}\cdot\nabla \psi({\bf x},t) +\lambda D\int_{-\infty}^t dt'
\exp(-\alpha\lambda(t-t')) \frac{\partial \psi({\bf x},t')}{\partial t'} = \nonumber \\ 
&-&\nabla^{-2}f'(z-h({\bf r},t))[\frac{\partial h({\bf r},t)}{\partial t} +{\bf v}\cdot\nabla h({\bf r},t)]  +\lambda D\int_{-\infty}^t dt'
\exp(-\alpha\lambda(t-t')) f'(z-h({\bf r},t'))\frac{\partial h({\bf r},t')}{\partial t'}\nonumber \\
&\approx& -\nabla^{-2}f'(z) [\frac{\partial h({\bf r},t)}{\partial t} +{\bf v}\cdot\nabla h({\bf r},t)] +\lambda D\int_{-\infty}^t dt'
\exp(-\alpha\lambda(t-t')) f'(z)\frac{\partial h({\bf r},t')}{\partial t'},\end{eqnarray}
where in the last line above we have neglected terms quadratic in $h$. 
Note that the neglecting of these additional terms is not strictly justified, they could potentially induce non-perturbative effects which render the surface fluctuations non-Gaussian. However we see here that the first order computation we carry out tends to reduce fluctuations with respect to equilibrium or non-driven interfaces and so if the equilibrium theory can be described by an equation which is linear in height fluctuations, it seems physically reasonable to assume that the the approximation also holds for the driven interface. 
Again, we multiply the above by $f'(z)$ and integrate over $z$. In the first term we make use of the approximation
\begin{equation}
f'(z)=\Delta\psi \delta(z)\label{eqdelta}
\end{equation}
and in the second we use the relation in Eq. (\ref{mfsig}). Putting this all together we obtain
\begin{equation}
\Delta\psi^2 \int d{\bf r} G(0,{\bf r}-{\bf r}') [\frac{\partial h({\bf r},t)}{\partial t} +{\bf v}\cdot\nabla h({\bf r},t)] +\frac{\sigma\lambda D}{\kappa}\int_{-\infty}^t dt'
\exp(-\alpha\lambda(t-t'))\frac{\partial h({\bf r},t')}{\partial t'}
= \sigma[\nabla^2 h({\bf r},t)-m^2 h({\bf r},t)] + \xi({\bf r},t),\label{em}
\end{equation}
where $G= -\nabla^{-2}$, or more explicitly
\begin{equation}
\nabla^2 G(z-z',{\bf r}-{\bf r}') = -\delta(z-z') \delta({\bf r}-{\bf r'}).
\end{equation}
The noise term $\xi$ is given by
\begin{equation}
\xi({\bf r},t) = \int_{-\infty}^{\infty} dz f'(z-h({\bf r},t)) \nabla^{-2} \zeta({\bf x},t).
\end{equation}
Now, as the equations of motion have been derived to first order in $h$ and we wish to recover the correct equilibrium statistics for the non-driven system, we ignore the $h$ dependence in the noise and make the approximation
\begin{equation}
\xi({\bf r},t) \approx \int_{-\infty}^{\infty} dz f'(z) \nabla^{-2} \zeta({\bf x},t).
\end{equation}
The correlation function of this noise is most easily evaluated in terms of its Fourier transform with respect to  space and time  defined by
\begin{equation}
\hat F({\bf q},\omega)=\int dt d{\bf r}\exp(-i\omega t -i{\bf q}\cdot{\bf r}) F({\bf r},t).
\end{equation}
Using the relations Eqs. (\ref{mfsig}) and (\ref{eqdelta}) one  can show that
\begin{equation}
\langle \hat \xi({\bf q},\omega)\hat \xi({\bf q}',\omega')\rangle 
=2T(2\pi)^d \delta(\omega +\omega') \delta({\bf q}+{\bf q}') \left[
\frac{\sigma}{\kappa}\frac{\alpha D^2\lambda^2}{\omega^2 +\alpha^2\lambda^2} + \frac{D\Delta\psi^2}{2q}\right].
\end{equation}
In full Fourier space the equation of motion for the field $\psi$ then reads
\begin{equation}
\left[i(\omega+{\bf q}\cdot{\bf v})\frac{\Delta\psi^2}{2q} + \frac{D\sigma\lambda}{\kappa} \frac{i\omega}{\alpha\lambda+i\omega}\right] \hat h({\bf q},\omega)= -D\sigma(q^2+m^2)\hat h({\bf q},\omega)+ \hat\xi({\bf q},\omega).\label{dyn}
\end{equation}

From this, the full Fourier transform of the correlation function of the interface height is given by
\begin{equation}
\hat C({\bf q},\omega)  = 2TD \frac{\left[ \frac{\Delta\psi^2}{2q}(\omega^2+\alpha^2 \lambda^2) + \frac{\sigma\alpha D\lambda^2}{\kappa}\right]}{\left|i[\frac{\alpha\lambda\Delta\psi^2}{2 q}(\omega + {\bf q}\cdot{\bf v}) + \frac{\lambda \sigma D}{\kappa}\omega + D\sigma(q^2+m^2)\omega]
+[\alpha\lambda D\sigma(q^2+m^2) -\frac{\Delta\psi^2}{2q}\omega(\omega+{\bf q}\cdot{\bf v})]\right|^2}.
\end{equation}
Using the above we can extract the equal time height-height correlation function in the steady states. Its spatial Fourier transform can shown to be given by
\begin{eqnarray}
\tilde C_s({\bf q}) &=& \frac{1}{2\pi} \int d\omega \hat C({\bf q}, \omega)\nonumber\\
&=&T \frac{\left(2 D\sigma q(\kappa[q^2+m^2]+\lambda)+\alpha\kappa\lambda\Delta\psi^2\right)^2 +\kappa^2 \Delta\psi^4 ({\bf q}\cdot{\bf v})^2}{\sigma[q^2+m^2]\left(2D q\sigma (\kappa[q^2+m^2]+\lambda)+\alpha \kappa\lambda \Delta\psi^2\right)^2 + \kappa\left(\kappa\sigma[q^2+m^2] + \lambda\sigma\right)\Delta\psi^4({\bf q}\cdot{\bf v})^2}.\label{eqmaind}
\end{eqnarray}
An outline of the derivation of this result is given in the Appendix to the paper.
In the absence of driving, {\em i.e.} when ${\bf v}={\bf 0}$ we recover the equilibrium correlation function
\begin{equation}
\tilde C_s({\bf q})= \tilde C_{eq}({\bf q})= \frac{T}{\sigma[q^2+m^2]},
\end{equation} 
here we see that  $1/m= \xi_{eq}$ is the so called capillary length, which is the equilibrium correlation length of the height fluctuations. We also notice that the correlation function for wave vectors perpendicular to the driving direction is simply the equilibrium one.

If we write $C_s({\bf q})= T/H_s({\bf q})$ we can interpret $H_s({\bf q})$ as an effective quadratic Hamiltonian for the height fluctuations, it is thus given by
\begin{equation}
H_s({\bf q}) = \sigma[q^2+m^2] + \frac{\kappa\lambda\sigma \Delta\psi^4 ({\bf q}\cdot{\bf v})^2}{\left(2 D\sigma q(\kappa[q^2+m^2]+\lambda)+\alpha\kappa\lambda\Delta\psi^2\right)^2 +\kappa^2 \Delta\psi^4 ({\bf q}\cdot{\bf v})^2}
\end{equation}
For small $q$ we find 
\begin{equation}
H_s({\bf q}) = \sigma m^2 + \sigma q^2(1+ \frac{v^2\cos^2(\theta)}{\alpha^2\lambda\kappa}),
\end{equation}
where $\theta$ is the angle between the wave vector ${\bf q}$ and the direction of the driving. 
This thus gives a direction dependent surface tension 
\begin{equation}
\sigma_s(\theta) = \sigma(1+ \frac{v^2\cos^2(\theta)}{v^2_0}),
\end{equation}
where we have introduced the intrinsic velocity $v_0 = \sqrt{\alpha^2\lambda\kappa}$ which depends on the microscopic {\em dynamical} quantity $\alpha$ associated with the model A dynamics of the field $\phi$, as well as the microscopic static quantities $\kappa$ (which generates the surface tension) and $\lambda$ the coupling between the field $\psi$ and $\phi$. This appearance of dynamical and static quantities that are otherwise hidden in equal time correlation functions in equilibrium is already implicit in the works of Onsager \cite{hem1996} where it is used to compute the conductivity of Brownian electrolytes and the explicit expressions were derived using stochastic density functional theory in \cite{dem2016}. We also note that the universal thermal Casimir effect between model Brownian electrolyte systems  driven by an electric field 
exhibits similar features, developing a dependency on both additional static and dynamical variables with respect to the equilibrium case \cite{dean2016}

However for this small $q$ expansion we see that the microscopic 
quantities $D$, the diffusion constant of the field $\phi$, and the order parameter jump
$\Delta\psi$ do not appear. 

From the above, we see that  in the direction of the driving the surface tension increases and the fluctuations of the surface are thus suppressed. We may also write 
\begin{equation}
H_s({\bf q}) = \sigma_s(\theta) [q^2 + m^2_e(\theta)],
\end{equation}
with 
\begin{equation}
m^2_s(\theta) =\frac{ m^2}{1+ \frac{v^2\cos^2(\theta)}{v_0^2}},
\end{equation}
this corresponds to a correlation length 
\begin{equation}
\xi_s = \xi_{eq}\sqrt{1+ \frac{v^2\cos^2(\theta)}{v_0^2}},
\end{equation}
and we see that it is increased in the direction of the driving. 

As we have just remarked  that the above results appear to be independent of the order parameter jump $\Delta \psi$ and the diffusion constant $D$, however the next order correction to $H_s$ for small $q$ is given by
\begin{equation}
H_s({\bf q}) = \sigma_s(\theta) [q^2 + m^2_e(\theta)] - \frac{4Dq \sigma^2(\lambda+\kappa m^2)( {\bf q}\cdot{\bf v})^2 }{\alpha^3 \kappa^2 \lambda^2 \Delta\psi^2},
\end{equation}
and so the small ${\bf q}$ expansion  breaks down at $\Delta\psi=0$, indeed one can see that the system has exactly the equilibrium correlation function when  $\Delta\psi=0$. 

In the limit of large $q$ we see that the effective Hamiltonian is given, to leading order, by the original equilibrium Hamiltonian and so the out of equilibrium driving has no effect on the most energetic modes of the system.

The results here predict that for unconfined surfaces the long range height fluctuations are described by an isotropic form of capillary wave theory with 
an anisotropic surface tension which is largest in the direction of driving. Numerical simulations of driven lattice gases in two dimensions \cite{leun1993} show a more drastic change upon driving and find $C_s(q)\sim  1/q^{.66}$ and thus a strong deviation from capillary wave theory.  
\section{A model of active interfaces}
We can apply the results derived in the previous section to analyse a simple model for
surfaces formed between two phases of active colloids. Activity is modelled by assuming that the colloidal field $\psi$ has a temperature different to that of  the solvent field $\phi$. This models the effect that activity leads to enhanced colloidal diffusivity over and
above the Brownian motion of particles due to thermal fluctuations \cite{gros2015}.

In the absence of any driving the dynamical equations for the field $\psi$ and $\phi$ become 
\begin{eqnarray}
\frac{\partial \psi({\bf x},t)}{\partial t} &=& D\nabla^2\frac{\delta H}{\delta \psi({\bf x})}+ \sqrt{2D T_1}\nabla \cdot {\bm \eta}_1({\bf x},t) \\
\frac{\partial \phi({\bf x},t)}{\partial t} &=& -\alpha\frac{\delta H}{\delta \phi({\bf x})}+ \sqrt{2\alpha T_2}{ \eta}_2({\bf x},t).
\end{eqnarray}
Following the same arguments as above we find that
\begin{equation}
\hat C({\bf q},\omega)  = 2D \frac{\left[ T_1\frac{\Delta\psi^2}{2q}(\omega^2+\alpha^2 \lambda^2) + T_2\frac{\sigma\alpha D\lambda^2}{\kappa}\right]}{\left|i\omega[\frac{\alpha\lambda\Delta\psi^2}{2 q} +  \frac{\lambda \sigma D}{\kappa} + D\sigma(q^2+m^2)]
+[\alpha\lambda D\sigma(q^2+m^2) -\frac{\Delta\psi^2}{2q}\omega^2]\right|^2}.
\end{equation}
The equal time steady state height fluctuations thus have correlation function
\begin{equation}
\tilde C_s(q) = \frac{T_1}{\sigma (q^2 + m^2)}\left[ 1 -(1-\frac{T_2}{T_1})\frac{\lambda\sigma } {\kappa }\frac{1}{\frac{\alpha\lambda \Delta \psi^2}{2Dq}+ \frac{\lambda\sigma }{\kappa} + \sigma(q^2+m^2)}\right].
\end{equation}
We see, again, that the inclusion of a non-equilibrium driving changes the statistics of height fluctuations and leads to a steady state that depends on both dynamical variables
$D$ and $\alpha$ as well as static ones $\Delta\psi,\ \lambda$ and $\kappa$ that remain hidden in the equilibrium case. This phenomenon is again seen in the behavior of the universal thermal  Casimir force between Brownian conductors held at different temperatures \cite{lu2015}.

If we assume strong activity we can take the limit $T_1\gg T_2$, in this case we find
\begin{equation}
\tilde C_s(q) = \frac{T_1}{\sigma (q^2 + m^2)}\frac{\frac{\alpha\lambda \Delta \psi^2}{2Dq}+
\sigma(q^2+m^2)}{\frac{\alpha\lambda \Delta \psi^2}{2Dq}+ \frac{\lambda\sigma }{\kappa} + \sigma(q^2+m^2)}.
\end{equation}
Interpreted in terms of an effective Hamiltonian for an equilibrium system at the temperature $T_1$ the above gives
\begin{equation}
H_s(q) = \sigma (q^2 + m^2)\left[1+\frac{\lambda\sigma }{\kappa}\frac{q}{\frac{\alpha\lambda \Delta \psi^2}{2D}+
q\sigma(q^2+m^2)}\right].
\end{equation}
In the case of an unconfined interface (where there is no gravitational effect
on the surface fluctuations) {\em i.e.} $m=0$ we see that for small $q$
\begin{equation}
H_s(q) \approx \sigma q^2 +\frac{2D\sigma^2 }{\kappa\alpha \Delta\psi^2}q^3 .
\end{equation}
We see that the effective surface tension is not modified but a reduction of fluctuations due to the presence of the term in $q^3$ arises.  As in the case of a driven system, we see that the large $q$ behavior of the effective Hamiltonian is given by the equilibrium case where $T=T_1=T_2$. 

In the case where the interface is confined, we see that for small $q$ one obtains
\begin{equation}
H_s(q) \approx \sigma m^2 \left[ 1+ \frac{2D\sigma }{\kappa\alpha \Delta\psi^2}q\right],
\end{equation}
and thus at the largest length scales of the problem there is a qualitative departure from capillary wave behavior, and the correlation length of height fluctuations at the largest length scales is given by
\begin{equation}
\xi_s = \frac{2D\sigma }{\kappa\alpha \Delta\psi^2}.
\end{equation}
The above result should be compared with that obtained in \cite{zia1991} for 
systems with anisotropic thermal white noise, which breaks detailed balance and mimics random driving of the system parallel to the interface; for free interfaces it was found that $C_s(q)\sim 1/q$.
\section{Conclusions}
We have presented a model to analyse the effect of uniform driving on the dynamics of the interface in a two phase system. In order to generate a non-equilibrium state a second {\em hidden} order parameter was introduced. This models the behaviour of a local or solvent degree of freedom which is not influenced by the driving field. In this way, we obtain out of equilibrium interface fluctuations which are described by Gaussian statistics as found in the experimental study of \cite{derks2006}. The agreement with this experimental study also extends to qualitative agreement with the increase of the effective surface tension in the direction of driving and also an increase in the correlation length of the height fluctuations with respect to a non-driven equilibrium interface. However, we  note that numerical simulations of a sheared Ising interface \cite{smith2008,smith2010} also reveal a reduction of interface fluctuations but the lateral correlation length is found to be reduced.

The basic idea underlying this study would be interesting to apply to a number of possible variants of this model, for instance both the dynamics
of the main field $\phi$ and the solvent field $\phi$ could be varied. To make a direct link with driven colloidal interfaces one should study model H type dynamics for the main field $\phi$ and other variants for the dynamics of the 
solvent field $\phi$ could also be considered. 

As mentioned above, in lattice based models driving induces non-equilibrium states even in the simple Ising lattice gas. A model analogous to that studied here can be formulated in a lattice based systems using the Hamiltonian 
\begin{equation}
H = -J\sum_{(ij)}S_i S_j (1+ \sigma_{(ij)}),
\end{equation}
where $S_i=\pm1$ are Ising spins at the lattice sites $i$, and $\sigma_{(ij)}=\pm 1$ are Ising like dynamical solvent variables associated with the lattice links $(ij)$. The static partition function is given by
\begin{equation}
Z = {\rm Tr}_{\sigma_{ij},S_i} \exp\left[\beta J\sum_{(ij)}S_i S_j (1+ \sigma_{(ij)})\right],
\end{equation}
and the trace over the solvent variables can be trivially carried out to give
\begin{equation}
Z = {\rm Tr}_{S_i}\left( \exp\left[\beta J\sum_{(ij)}S_i S_j \right]\prod_{(ij)}2\cosh(\beta JS_iS_j)\right )= [2\cosh(\beta J)]^L{\rm Tr}_{S_i}\exp(\beta J\sum_{(ij)}S_i S_j ),
\end{equation}
where $L$ is the number of links on the lattice of the model. We thus see that the underlying effective static model is precisely the zero field Ising model. 

This model can then be driven in a number of ways, for instance using conserved Kawasaki dynamics for the Ising spins to model diffusive dynamics in the presence of a uniform driving field parallel to the surface between the two phases at a temperature below the ferromagnetic ordering temperature $T_c$. The dynamics of the Ising spins on the lattice links can  be given by non-conservative single spin flip, for instance Glauber, dynamics to keep the analogy with the continuum model discussed in the paper but diffusive dynamics or indeed a mixture of diffusive and non-conserved dynamics 
could be implemented. It would be interesting to see to what extent this modification of the driven lattice gas model affects the non-equilibrium driven states that arise. 

It is also clear that this lattice model can be used to simulate the effect of activity where the Ising spins $S_1$ corresponding to the colloid field undergo  Kawasaki dynamics at the temperature $T_1$ where as the link variables $\sigma_{(ij)}$ undergo single spin flip non-conserved dynamics
at the temperature $T_2$.

\section{Acknowledgements}
The authors acknowledge support from the ANR (France) Grant FISICS \\
\appendix
\section{Evaluating Fourier integrals}
Here we outline how the Fourier integration leading to Eq. (\ref{eqmaind}) is carried out. Defining
\begin{equation}
I(f(\omega)) = \int \frac{d\omega}{2\pi} \frac{f(\omega)}{\left|i(A\omega + B) + (C-D\omega-E \omega^2)\right|}
\end{equation}
we see that the integral we need to evaluate can be written in the form
\begin{equation}
I = a I(\omega^2) + b I(1).
\end{equation}
The calculation leading to Eq. (\ref{dyn}) can be carried out in the presence of a forcing term on the height profile in order to compute the response function for the surface which has a denominator of the form
\begin{equation}
{\rm Den} = i(A\omega + B) + (C-D\omega-E \omega^2),
\end{equation}
and due to causality the above only has poles in the upper complex plane (due to the convention of Fourier transforms used here). Consequently we find that
\begin{equation}
\int \frac{d\omega}{2\pi} \frac{1}{i(A\omega + B) + (C-D\omega-E \omega^2)} = 0,\label{key}
\end{equation}
as one may close the integration contour in the lower half of the complex plane. Taking the real and imaginary part of Eq. (\ref{key}) leads to
\begin{eqnarray}
C I(1) -D I(\omega) - E I(\omega^2) = 0 \\
AI(\omega) + B I(1) = 0.
\end{eqnarray}
Using this we can express $I(\omega^2)$ as a function of $I(1)$, and explicitly we have 
\begin{equation}
I(\omega^2) = \frac{I(1)}{E}[C+ \frac{DB}{A}].
\end{equation}

To evaluate $I(1)$ we now use
\begin{equation}
I(1) = -{\rm Im} \int \frac{d\omega}{2\pi}\frac{1}{A\omega +B} \frac{1}{i(A\omega + B) + (C-D\omega-E \omega^2)}.
\end{equation}
The integrand above has no poles in the lower half of the complex plane but has a {\em half pole} at $\omega=-B/A$ on the real axis, thus using standard complex analysis we find
\begin{equation}
I(1) = \frac{1}{2(CA + BD - \frac{EB^2}{A})}.
\end{equation}
Then after some laborious, but straightforward algebra, the results Eq. (\ref{eqmaind}) is obtained.


\begin{thebibliography}{3}
\bibitem{derks2006} D. Derks, D. G. A. L. Aarts, D. Bonn, H. N. W. Lekkerkerker,
and A. Imhof, Phys. Rev. Lett. {\bf 97}, 038301 (2006).
\bibitem{smith2008}T. H. R. Smith, O. Vasilyev, D. B. Abraham, A. Maciolek, and M. Schmidt, Phys. Rev. Lett. {\bf 101}, 067203  (2008)
\bibitem{smith2010}T. H. R. Smith, O. Vasilyev, A. Maciolek, and M. Schmidt
Europhys. Lett. {\bf 89} 10006 (2010).
\bibitem{hem1996}
P.C. Hemmer, H. Holden and S. Kjelstrup~Ratkje,   {\em {The collected works of Lars Onsager: with commentary}\/} vol~17 ({Singapore}: {World Scientific}) (1996).
\bibitem{netz2003}R.R. Netz, Europhys. Lett. 63, 616 (2003).
\bibitem{dzub2002}J. Dzubiella, G.P. Hoffmann, and H. L\"ošwen, Phys. Rev. E {\bf 65}, 021402 (2002).
\bibitem{chak2003} J. Chakrabarti, J. Dzubiella, and H. Lš\"owen, Europhys. Lett. {\bf 61}, 415 (2003).
\bibitem{chak2004} J. Chakrabarti, J. Dzubiella, and H. Lš\"owen,  Phys. Rev. E {\bf 70}, 012401 (2004).
\bibitem{lowe2009} H. Lš\"owen, Phys. Rev. Lett. {\bf 102}, 085003 (2009).
\bibitem{glan2012} T. Glanz and H. L\"ošwen, J. Phys. Condens. Matter {\bf 24}, 464114 (2012).
\bibitem{klym2016} K. Klymko, P.L. Geissler, and S. Whitelam,  Phys. Rev. E {\bf 94}, 022608 (2016).
\bibitem{dem2016}V. D\'emery and D.S. Dean, J. Stat. Mech. 023106 (2016).
\bibitem{pon2017}A. Poncelet, O. B\'enichou, V. D\'emery and G. Oshanin, Phys. Rev. Lett. {\bf 118}, 118002 (2017)
\bibitem{dean1996} D.S. Dean, J. Phys. A {\bf 29}, L613 (1996).
\bibitem{katz1984} S. Katz, J.L. Lebowitz and H. Spohn, J. Stat. Phys. {\bf 34}, 497 (1984).
\bibitem{zia1991}  R.K.P. Zia and K.-t. Leung, J. Phys. A Math. Gen. {\bf 24}  L1399 (1991).
\bibitem{leun1993} K.-t. Leung and R. K. P. Zia, J. Phys. A {\bf 26}, L737 (1993).
\bibitem{schm1995} B. Schmittmann and R. K. P. Zia, Phase Transitions and Critical Phenomena, Vol. 17, edited by C. Domb and J. L. Lebowitz, Academic. London (1995).
\bibitem{schm1998}B. Schmittmann and R. K. P Zia, Phys. Rep. {\bf 301}, 45 (1998).
\bibitem{bray2001}A.J. Bray, A. Cavagna and R.D.M. Travasso, Phys. Rev. E {\bf 64}, 
012102 (2001).
\bibitem{bray2002}A.J. Bray, A. Cavagna and R.D.M. Travasso, Phys. Rev. E {\bf 65}, 
016104 (2002).
\bibitem{thie2010} M. Thi\'ebaud and T. Bickel, Phys. Rev. E {\bf 81}, 031602 (2010).
\bibitem{thie2014}M.Thi\'ebaud, Y. Amarouchene, and T. Bickel
J. Stat. Mech.  P12011 (2014).
\bibitem{hohe1977}P.C. Hohenberg and B.I. Halperin,  Rev. Mod. Phys. {\bf 49}, 435 (1977).
\bibitem{dean2016}D.S. Dean, B.-S. Lu, A.C. Maggs and R. Podgornik, Phys. Rev Lett. {\bf 116}, 240602 (2016).
\bibitem{gros2015}A. Y. Grosberg and  J.-F. Joanny, Phys. Rev. E {\bf 92}, 032118 (2015).
\bibitem{lu2015}B.-S. Lu, D.S. Dean, and  R. Podgornik, {\em Europhys. Lett.} {\bf 112}, 20001 (2015).
\end{thebibliography}
\end{document}